# How a Supercooled Liquid Borrows Structure from the Crystal

Ulf R. Pedersen[1], Ian Douglass[1,2] and Peter Harrowell[2,*]

[1] *Glass and Time, IMFUFA, Department of Science and Environment, Roskilde University, P.O. Box 260, DK-4000 Roskilde, Denmark*

[2] *School of Chemistry, University of Sydney, Sydney New South Wales 2006 Australia*

* corresponding author  peter.harrowell@sydney.edu.au

Abstract

Using computer simulations, we establish that the structure of a supercooled binary atomic liquid mixture consists of common neighbour structures similar to those found in the equilibrium crystal phase, a Laves structure. Despite the large accumulation of crystal-like structure, we establish that the supercooled liquid represents a true metastable liquid and that liquid can 'borrow' crystal structure without being destabilized. We consider whether this feature might be the origin of all instances of liquids of a strongly favoured local structure.

The notion that glasses are rigid due the aperiodic accumulation of low energy local structures has had some success. This idea is the basis of Zachariasen's model of amorphous silica based on a random network of (slightly distorted) $SiO_4$ tetrahedra [1]. The validity of this model of silica has been supported by experiment [2] and simulation [3]. Water also accumulates, on supercooling, local tetrahedral coordination of hydrogen about each oxygen to form the Low Density Amorphous phase [4]. In each of these cases, the low energy structure is very similar to that found in the stable crystal phase. While crystal-like structures have also been invoked to describe supercooled atomic mixtures [5], the majority of studies of glass-forming alloys have focussed on non-crystalline local structures [6,7]. Over the years, a range of candidate structures have been proposed, starting with Franck's suggestion [8] of the regular icosahedron. The difficulty with this type of analysis is the large structural diversity typically found in atomic mixtures [9] and the associated low concentration of nominated local structures of interest. There are, however, exceptions to this scarcity of favoured local structures. In 2007, Coslovich and Pastore [10] demonstrated a substantial accumulation of local icosahedral structures in a supercooled equimolar binary Lennard-Jones mixture introduced by Wahnström [11]. As a result, the Wahnström liquid has become an important exemplar of the proposition that supercooled liquids can be characterised by a favoured local structure [6,7].

In 2010 it was shown [12] that the Wahnström liquid freezes into a crystal with the $MgZn_2$ (C14) Laves structure at a composition $A_2B$ (see Fig.1a). In this crystal, along with the related $Cu_2Mg$ (C15) Laves structure, *every* small particle (i.e. the A particles) has an icosahedral coordination shell. In this sense, the icosahedra in this supercooled liquid can be regarded as a structure borrowed from the crystal. Since Laves structures are the most abundant crystal forms found in binary intermetallics [13], this structural link between icosahedra in supercooled liquids and the equilibrium crystal phase is likely to be relevant to a wide range of metal alloys.

When we observe crystal-like structure in a supercooled liquid is the observed structure an intrinsic feature of the liquid or is it a result of an instability of the liquid, possibly arrested, with respect to crystallization? Moore and Molinero [14] reported that a model of water based on modified silicon potential exhibited an instability to crystallization at a temperature close to that associated with the onset of local ordering. No such instability has been observed in more complex models of water [15]. Fitzner et al [16] have, however, reported the presence of 6-fold rings of hydrogen bonded molecules (a vestige of the equilibrium crystal phase) in supercooled water that are associated with preferred sites of crystal nucleation. The goal of this paper is to establish i) the degree to which the local structure of the supercooled Wahnström liquid shares the common neighbour organization found in the crystal, and ii) to determine whether this structure is an intrinsic feature of the metastable liquid or is it the result of an incipient instability of the liquid with respect to crystallization?

The Wahnström mixture [11] consists of particles interact via Lennard-Jones potentials $\phi_{ij}(r) = 4\varepsilon_{ij}\left[(\sigma_{ij}/r)^{12} - (\sigma_{ij}/r)^{6}\right]$ with the following interaction parameters: $\sigma_{AA} = \sigma_{AB}/1.1 = \sigma_{BB}/1.2$ and $\varepsilon_{AA} = \varepsilon_{AB} = \varepsilon_{BB}$. The particle masses are related via $2m_A = m_B$. As we are interested in exploring the possible instability of the liquid with respect to growth of the $A_2B$ crystal, we have chosen to study the liquid, not at the equimolar concentration usually used in this model, but the $A_2B$ concentration, congruent with that of the crystal. All calculations have been carried out at a fixed reduced density of 0.874. At this density, the freezing point lies in the range $0.94 \leq T_f \leq 1.1$. Molecular dynamic (MD) simulations have been carried out at fixed NVT using RUMD [17] with N = 576 (unless otherwise indicated) using the Nose-Hoover thermostat [18].

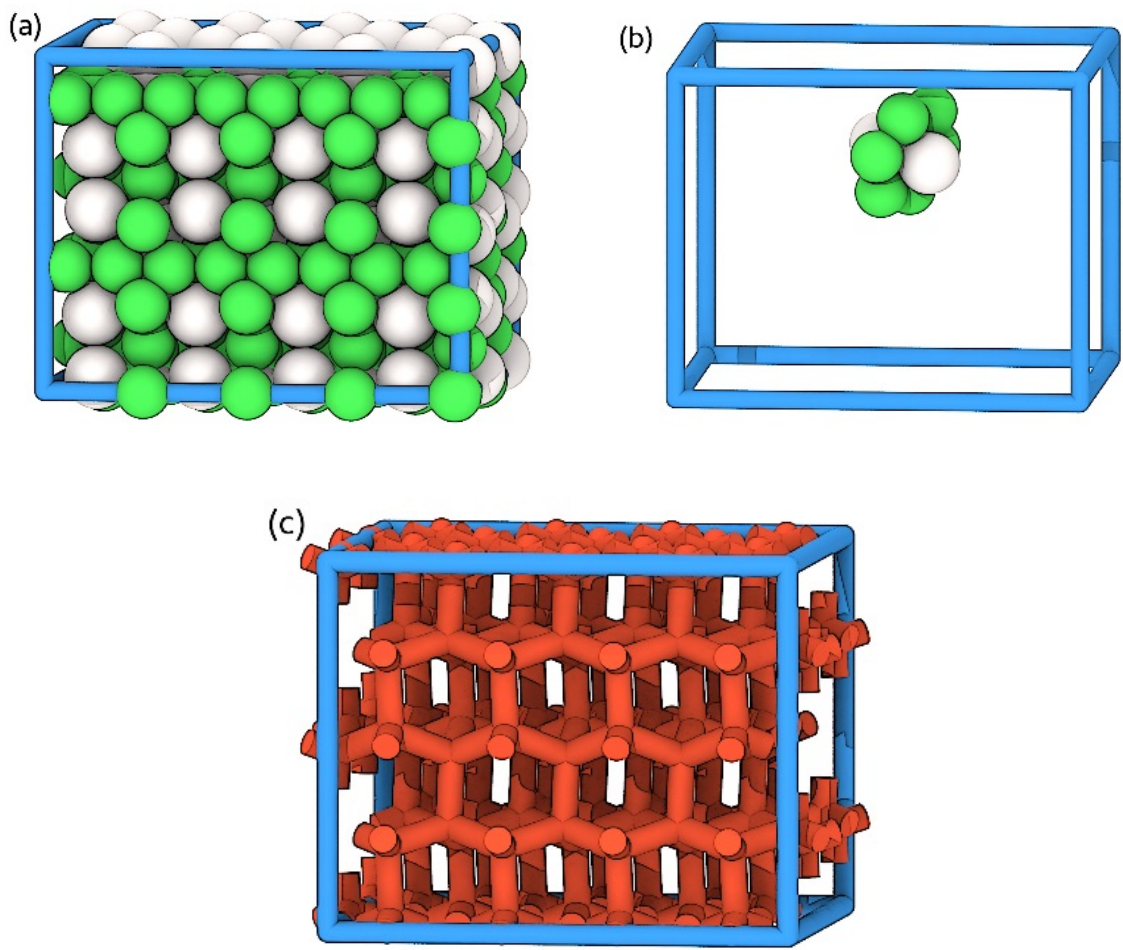

**Figure 1.** a) Illustration of the MgZn$_2$ crystal structure with A and B particles shown as green and grey, respectively. b) A single pair of B particles with their 6 common A neighbours from the crystal corresponding to a Frank-Kasper (FK) bond. c) The MgZn$_2$ crystal depicted as a tetrahedral network of FK bonds.

The MgZn$_2$ crystal (see Fig. 1a) can be represented in terms of 'bonds' between neighbouring pairs of B particles, where each bond consists of 6 common A neighbours to the two 'connected' B particles as shown in Fig.1b. (Two particles are identified as neighbours if their separation is less that the first minimum of the species-appropriate pair distribution function .) We have previously called these common neighbour configurations Frank-Kasper (FK) bonds [12]. The $MgZn_2$ crystal can be represented as a hexagonal diamond network of FK bonds in which each B particle occupies a 4-fold vertex (see Fig.1c) and every A particle at the centre of an icosahedron consisting of six A and six B particles. This is a hierarchical structure with well-defined short-range order (the icosahedral coordination) and intermediate range order (the FK bonds). The intermediate structure can be resolved using two distinct order parameters: one, a measure of crystallinity $X$ ($= N_{4FK}/N_B$) where $N_{4FK}$ is the number of B particles participating in 4 FK bonds, and, a second order parameter, the mole fraction of FK bonds $n$ ($= N_{FK}/N_B$) where $N_{FK}$ is the number of FK bonds. This two-order parameter description encompasses the possibility that the liquid could accumulate FK bonds *without* accumulating the 4-fold vertices characteristic of the crystal. A liquid might, in this way, 'borrow' stable FK bond structures from the crystal while eschewing crystallinity itself.

To understand the stability of the liquid as a function of the density of FK bonds we need to be able to control the latter quantity. In the metastable liquid this is complicated by the background crystallization rate. A biased Monte Carlo (BMC) algorithm allows us to control the value of $n$ in the liquid at a temperature high enough that the intrinsic nucleation kinetics cannot interfere. The biased Monte Carlo (BMC) calculations with a harmonic weighting for a target value of $n = n_o$ [19] at temperatures in the range $0.6 \leq T \leq 0.8$. (Note that the actual mean value $n$ of the FK bonds achieved by the BMC is typically less than the target $n_o$ and we shall refer to the actual value of $n$ throughout this paper.) These calculations were carried out at the $A_2B$ composition with $N = 576$ and at fixed density of 0.874. The BMC method will allow us to determine the degree to which the liquid can accumulate FK bonds before being forced into crystallinity.

We shall work with the reduced structural landscape spanned by the space of $(n,X)$ depicted in Fig. 2. This plane includes regions that are impossible to achieve. It is, for example, not possible to have more 4-fold FK sites than there are FK bonds to make them. It is equally impossible to avoid 4-fold sites once the FK bond concentration is high enough. These inaccessible regions of the *(n,X)* plane are indicated in Fig. 2 by shading. The red curve represents to the relationship between n and X, $X = \left(\frac{n}{2}\right)^4$, corresponding to the case where the bonds are distributed randomly of bonds with no preference to aggregate into 4-fold crystal vertices. The straight line (green) represents the other limit, i.e. the case where the *only* FK bonds present are those taking part in crystal vertices. The percolation line (dashed) represents an interpolation of the random bond percolation on the diamond lattice [20] (at n = 0.86) and the site percolation on the same lattice (at n = 0.78) [21]. The values of n and X from the BMC calculations are plotted in Fig. 2. We find the simulation structures lie quite close to the line predicted assuming a random distribution of FK bonds; one in which the crystalline organization in the form of the 4FK sites occur only as required by random aggregation. We shall refer to this observed liquid structure as a *quasi-random* distribution of FK bonds. The quasi-random structure of bonds includes an accompanying concentration of icosahedral coordination shells about the small particles. As previous studies [10], have analysed the amorphous structures in terms of local icosahedral coordination, rather than FK bonds, it is useful to establish the correlation between the two quantities. In Fig. 3 we plot the fraction $f_{icos}$ (= $N_{icos}/N_A$) of A particles in icosahedral environments as a function the FK bond density. The data comes from weighted simulations of a high temperature (T = 0.9) liquid where no sign of crystallinity (i.e. 4-fold FK sites) was observed. The observed increase in the fraction of icosahedra due to the increase in the FK bond count at a fixed high temperature corresponds to ~ 0.5 of the A particles involved in FK bonds being in icosahedral environments. We conclude that the presence of FK bonds induces icosahedral ordering, but with a degree of correlation of roughly 2 icosahedra per FK bond in the quasi-random distribution, well below the value of 6 icosahedra per FK bond found in the ordered $MgZn_2$ crystal.

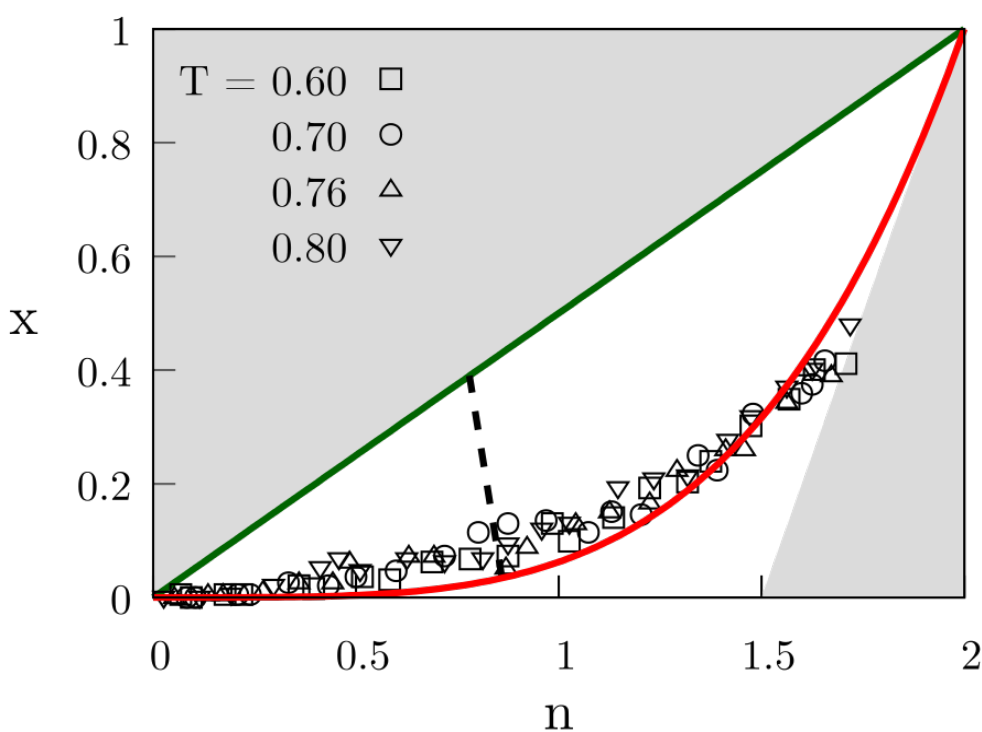

**Figure 2.** Plot of X, the crystalline order parameter, against n, the fraction of FK bonds, obtained from the biased MC calculations of the $A_2B$ mixture at the different temperatures shown. At the density of 0.874, the freezing point lies in the range $0.94 \leq T \leq 1.1$ [10]. The predictions for the randomly distributed bonds and for the case where bond occur only in crystal are shown as a red curve and a green line, respectively. The estimated percolation transition is shown as a dashed line and shaded areas indicate unphysical parts of the (n,X) plane (see text).

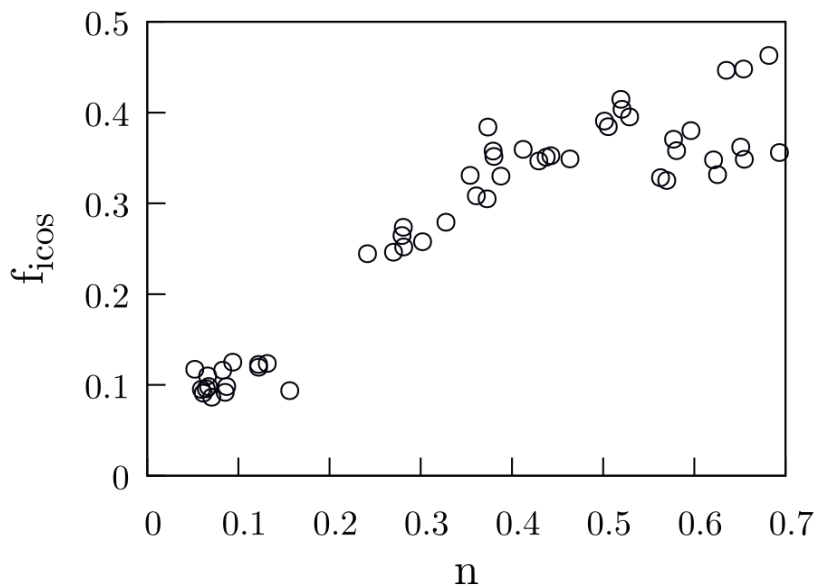

**Figure 3.** The fraction $f_{icos} = N_{icos}/N_A$ of icosahedral coordination shells as a function of the fraction $n = N_{FK}/N_B$ of FK bonds in the $A_2B$ mixture. The data comes from weighted simulations of a high temperature (T = 0.9) liquid where no sign of crystallinity (i.e. 4-fold FK sites) was observed.

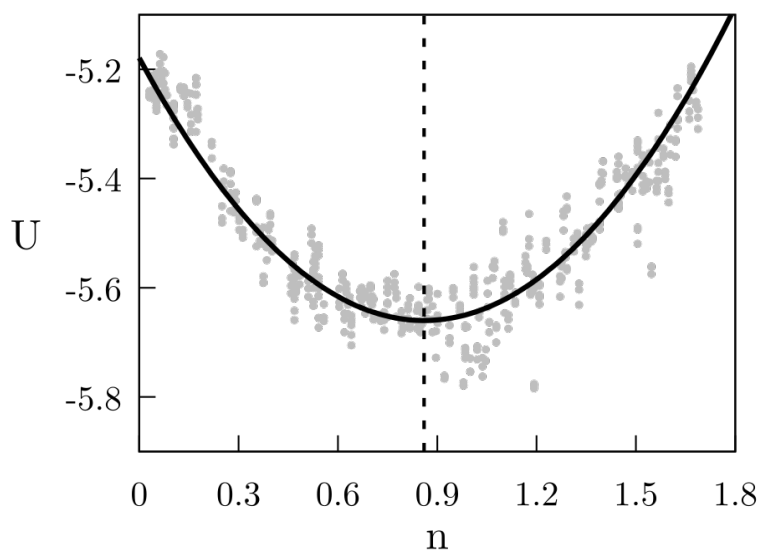

**Figure 4.** The energy per particle, U, of configurations with different values of n. All calculations were carried out at T = 1.0. The vertical dashed line indicates the random bond

percolation value of n= 0.86 [20]. The curve corresponds to a quadratic fit $U - U_o = c(n - 0.86)^2$ where $U_o$ = -5.66 and c = 0.65.

What is the energetics of this quasi-random assembly of FK bonds? In Fig. 4 we have calculated the potential energy per particle of the liquid, generated via the BMC method, as a function of *n*, the density of FK bonds. We find a clear potential energy minimum at *n* ~ 0.9. This corresponds closely to the value *n* = 0.86 for bond percolation on a diamond lattice, suggesting that the FK bonds added beyond the random percolation point introduce strain into the liquid structure. The increase in U for *n* > 0.9 via the continuation of the quasi-random distribution of FK bonds means that this structural option is no longer favoured on cooling. The fact that we generate these high energy states using BMC suggests that the algorithm only samples a restricted portion of configurations space, most probably due to the applied bias weighing against reorganizations that proceed via intermediates with low *n*. Any path to lower energy states must, therefore, involve deviations from the quasi-random structure, deviations characterised by an increase in the crystalline order parameter X. The value of *n* at percolation, therefore, represents an upper bound on the value at which this deviation will occur.

To find out how (n,X) order actually accumulates we have carried two sets of MD calculations. In Set I, a liquid was instantly quenched to the final temperature T and then annealed for a time of $10^2\tau_s(T)$ (where $\tau_s(T)$ is the structural relaxation time at T). In Set II, the annealing time was fixed, for all values of T, at $\sim 3 \times 10^7 \tau$, a value greater than $10^2\tau_s(T)$ for T > 0.63. The results for the potential energy U and the fraction of FK bonds *n* vs T for the two sets of simulations are plotted in Fig. 5a and 5b, respectively. For Set I, we see no sign of crystallization. The potential energy U decreases linearly with T, dropping well below the minimum value found from the BMC calculations. As shown in Fig. 5b, the accumulation of FK bonds in Set I as it reaches the glass state does not exceed a value $n = 0.2$. For Set II, we observe a step discontinuity in U and n at T = 0.89 which we attribute to crystallization. Increasing the system size by an order of magnitude increases the crystallization point to T = 0.92, but does not otherwise change the form of the data [22]. As we quench to temperatures below this freezing point, we observe that the value of n decreases, indicating a decreasing degree of crystallization, either through the decrease in crystallite size or the trapping of an increasing degree of disorder within crystallites. This decrease in FK bonds continues until n coincides with that of the glasses generated by the Set I runs. Despite the decrease in FK bonds, the energy U shows little variation for the Set II quenches below the freezing point.

The summary of our findings, presented graphically in Fig. 6, are as follows. 1) Under the restricted sampling of the BMC algorithm, we have established the existence of a *possible* liquid structure comprised of a quasi-random network of FK bonds with an energy minimum associated with the bond percolation point. 2) The arrested disordered structure, as generated by the Set I quench, consists of a quasi-random arrangement of FK bonds, but one with a maximum FK bond density of only $n = 0.2$ (or $f_{icos} \sim 0.25$-$0.3$). This value is well short of the percolation value as indicated in Fig. 2, leaving us to conclude that percolation plays no role in capping the accumulation of FK bonds in the metastable liquid. 3) The crystallization of the liquid is discontinuous and associated with substantial increases in n (and X). The structure clearly deviates from the quasi-random model. The abrupt and large increase in n associated with crystallization is quite distinct from the constrained accumulation of FK bonds (bounded by the value $n = 0.2$) that we associate with the supercooled liquid state and the glass that it becomes at low T. This is clear evidence that the FK bonds found in the arrested state correspond to the intrinsic structure of the metastable liquid and are not associated with any instability with respect to crystallization.

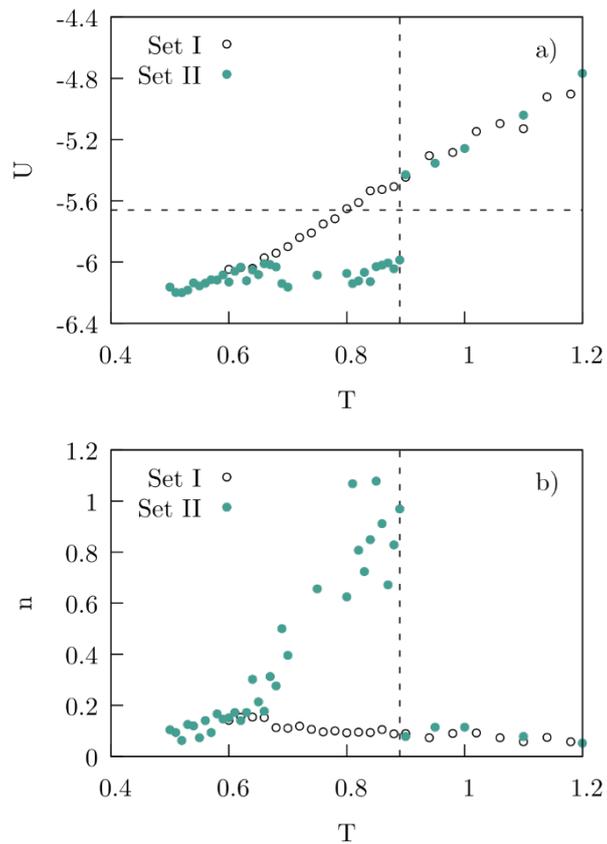

**Figure 5.** a) The average potential energy as a function of T. The results of MD simulations for Set I and II (see text) are shown. A horizonal dashed line is included to indicate the value of the potential energy at the minimum in Fig. 4, and a vertical dashed line to indicate the crystallization temperature T = 0.89. b) The average value of the mole fraction of FK bonds n as a function of T with data from MD simulations for Set I and II.

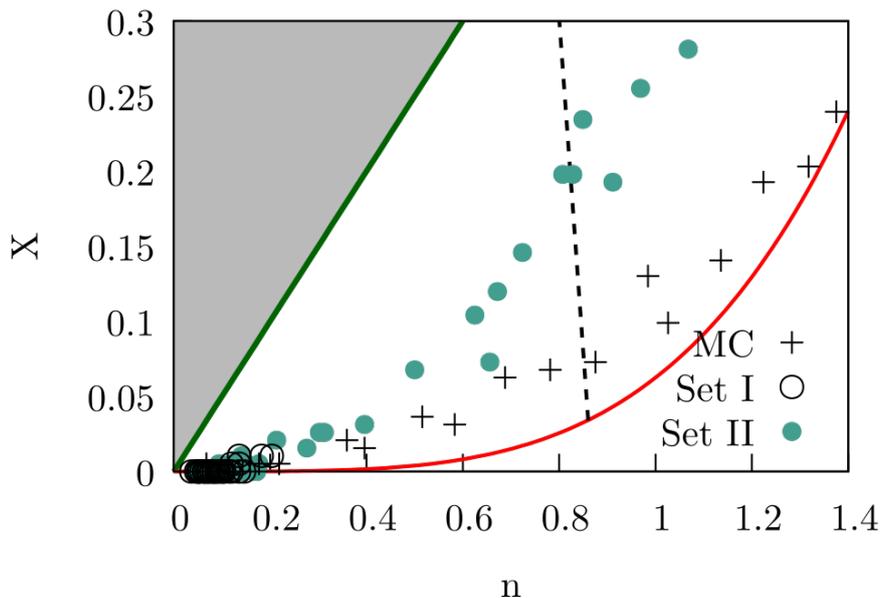

**Figure 6.** The (n,X) structure landscape including the results for the MD simulations of Set I and Set II (see text). Also plotted are the results of the BMC calculations, the percolation line and the two limits as presented in Fig. 2.

In conclusion, the structure of the arrested amorphous state of the $A_2B$ Wahnström mixture after the $10^2\tau_s$ annealing time (i.e. Set I calculations) consists of a significant fraction of common neighbour structures borrowed from the crystal phase. The value of n appears to be capped so that it does not increase continuously to the point of destabilizing the liquid. Having dismissed the bond percolation as the origin of this cap, we conclude that the most likely constraint is the slowing kinetics associated with the quasi-random ordering. The convergence of the low T structures from the Set I and Set II runs supports this kinetic explanation. Derlet and Maaß [23] have reported that, in extended 80 μs molecular dynamics simulations, the supercooled Wahnström model formed a nano-structured composite of the $A_2B$ Laves crystal and B-particle rich glass. Our observation of a discontinuous ordering to some partially crystalline state is consistent with this recent report.

The network liquids like silica and water have long represented the best known examples where crystal-like structure is observed in the liquid. The object of this paper is to understand how the analogous process is manifest in dense packed liquids. The 'borrowing' of crystal structure by the liquid that we observe in the Wahnström mixture is not a universal feature of alloys, with many liquid alloys failing to exhibit any sign of local crystal-like features [6,7]. As an example of this, the Kob-Andersen mixture [24], another binary Lennard-Jones mixture studied in the context of glass formation, exhibits favoured local structures quite distinct from those that appear in the crystal [25]. What is special about the model mixture studied here is its crystal structure. The Laves forming alloys are characterised by a low energy short range order (the icosahedra) that can readily organize into an intermediate range structure (the FK bond). This kind of hierarchical ordering lends itself to the stable incorporation of crystal-like structure in the liquid. It is likely that this structural borrowing may be a generic feature of liquid mixtures that freeze into a Laves phase or, possibly, for which the Laves phase is a metastable polymorph. This proposition is consistent with the experimental of the coincidence of glass forming ability and the stability of a Laves phases in a range of Fe-based alloys [26]. It would be interesting to test the generality of the connection, proposed here, between Laves crystals and the stability of supercooled alloys. An open question is whether this 'borrowing' of structure might be a general feature whenever the equilibrium crystal has a large unit cell. In establishing how fragments of complex crystal structures can play an important role in the structure of the supercooled liquid, this work points to a potentially useful new approach to identifying classes of amorphous materials assembled from well characterised local configurations.

**Supplementary Material**

The Supplementary Material provides additional data related to the influence of system size on the plots presented in Figs. 5a and 5b.

**Acknowledgements**

URP and ID acknowledge support from the VILLUM Foundation's *Matter* grant (No. 16515). ID and PH acknowledge funding support from the Australian Research Council.

**Data Availability**

The data that support the findings of this study are available from the corresponding author upon reasonable request.

Supplementary Material

**Role of System Size on Structure and Potential Energy**

In Fig. S1 we plot the values of the potential energy U and the fraction of FK bonds n as a function of T for the two cooling protocols, Set I and Set II, described in the text for two systems sizes, N = 576 and 5760. We find a close similarity between the values for the small and large systems. The obvious point of difference is that the freezing transition is shifted from T = 0.89 in the small system to T = 0.92 in the larger system, as indicated in the main text.

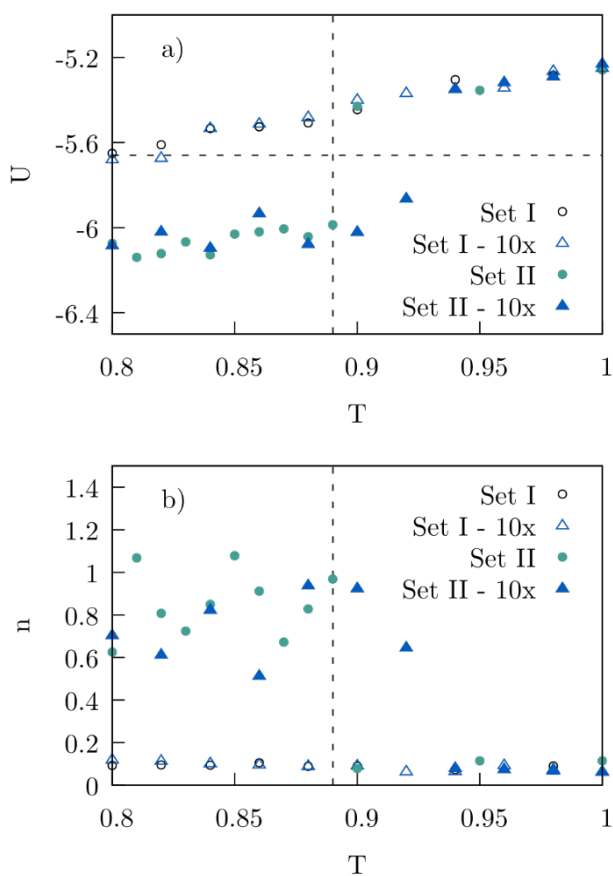

**Figure S1.** a) The potential energy and b) the fraction n of FK bonds as a function of temperature T for MD simulations using the two different cooling protocols, Set I and Set II, described in the main text. Data for two different values of N are shown, with the N = 576 data shown as circles and the N = 576 data shown as triangles. Set I and Set II data are indicated by open and filled symbols, respectively.